# Threshold Verification Technique for Network Intrusion Detection System

Faizal M. A., Mohd Zaki M., Shahrin S., Robiah Y, Siti Rahayu S., Nazrulazhar B.
Faculty of Information Technology and Communication
Univeristi Teknikal Malaysia Melaka,
Ayer Keroh, Melaka,
Malaysia
faizalabdollah@utem.edu.my, zaki.masud@utem.edu.my, shahrinsahib@utem.edu.my, robiah@utem.edu.my,
sitirahayu@utem.edu.my, nazrulazhar@utem.edu.my

Internet has played a vital role in this modern world, the possibilities and opportunities offered are limitless. Despite all the hype, Internet services are liable to intrusion attack that could tamper the confidentiality and integrity of important information. An attack started with gathering the information of the attack target, this gathering of information activity can be done as either fast or slow attack. The defensive measure network administrator can take to overcome this liability is by introducing Intrusion Detection Systems (IDSs) in their network. IDS have the capabilities to analyze the network traffic and recognize incoming and on-going intrusion. Unfortunately the combination of both modules in real time network traffic slowed down the detection process. In real time network, early detection of fast attack can prevent any further attack and reduce the unauthorized access on the targeted machine. The suitable set of feature selection and the correct threshold value, add an extra advantage for IDS to detect anomalies in the network. Therefore this paper discusses a new technique for selecting static threshold value from a minimum standard features in detecting fast attack from the victim perspective. In order to increase the confidence of the threshold value the result is verified using Statistical Process Control (SPC). The implementation of this approach shows that the threshold selected is suitable for identifying the fast attack in real time.

Keyword; fast attack detection, Intrusion detection system, Statistical Process Control

I. INTRODUCTION

Incidents of cyber attack on the Internet are increasing every year [1] and most administrators are depending on Intrusion Detection System (IDS) as the essential component in protecting their network. These attacks are generated using tools and exploits script which are freely available on the internet. Mc Hugh also provide further evidence by stating that anyone can attack Internet site using readily made available intrusion tools and exploit script that capitalize on widely known vulnerabilities [2]. Hence, the increasing number of the exploit tools may have influence on the number of novice attackers on the internet [3].

An attack can be divided into five phases which are reconnaissance, scanning, gaining access, maintaining access and covering tracks [4]. The first two phases is an initial stage of an attack and it does involve scanning and probing network traffic for information on the vulnerabilities of the targeted machine. This initial stage of attack can be categorized into two which are fast attack and slow attack; [5] defined fast attack as an attack that uses a large amount of packet or connection within a few second. Meanwhile a slow attack is defined as an attack that takes a few minutes or a few hours to complete [6]. This research focuses on fast attack scenario as this can prevent any early attack and may help to reduce the possibilities of further attack on the organization network.

In this research we are focusing on detecting fast attack based on the connection made by attacker on a single victim. This early detection could help the administrator to take action in preventing the next phase of the attack and investigate the reason and how the attack happened. The normal and the abnormal traffic are differentiated using a threshold value that is acquired from the result of the observation and experimental technique applied in the implementation of this research. To make it more reliable, the threshold value is verified using statistical control process approach.

The rest of this paper, we will discuss the background of this research in section 2. Section 3 will concentrate on methodology and techniques used in producing the threshold value and section 4 discuss the implementation and result of the research. Finally section 5 conclude and discuss the direction of this research.

II. BACKGROUND

Currently, most researcher concentrate on the individual attack such as DDOS [7][8], Worm [9][10], portscanning

[11] rather than intrusive behavior. [12] suggested the intrusion detection should focus on behavior rather that individual attack and by looking the intrusive behavior, it might offer an opportunity to increase the detection of intrusion [13]. Therefore, motivated by this opportunity, we classified the first two phases of anatomy of attack into two types which are fast attack and slow attack. The definition of fast attack and slow attack has been stated earlier. This research focuses on fast attack detection rather than slow attack due to prevention of early attack on the network may help to reduce the possibilities of gaining access to the vulnerable machine.

*A. Framework*

[14] has introduced a fast attack framework with a minimum set of feature selection to detect a fast attack at the early stage.

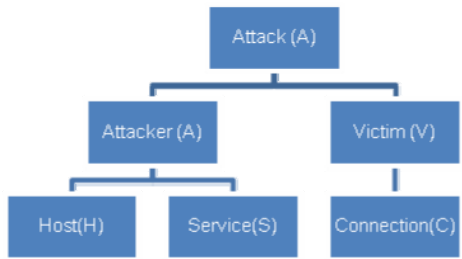

Figure 1. Fast attack framework

The framework classified attack into 2 perspectives, the attacker and victim. The attacker perspective deal with attack that is cause by a single host and to make it more reliable the attacker perspective can be further divided into 2 subcategories which are the attack that originated from a single host towards multiple hosts and the attack that originated from single host towards the services offered by the host such as scanning attack.

TABLE I. Feature selection

| Feature | Description | Category |
|---|---|---|
| *Timestamp* | Time the packet was send | |
| *Duration* | Duration of connection. | |
| *IP* | Addresses of host | |
| *Protocol* | Connection protocol (e.g tcp) | |
| *Flag* | Status flag of the connection | |
| *Service* | Source and Destination services. | |
| *Dest_count* | Number of connection having the same destination host (AVC). | Derived Features |

Meanwhile the victim perspective is focuses more on number of attackers connected to a single host. This research focuses on the victim perspective so that any initial attack on the victim from any attacker can be detected.

For the selection features [14] integrate a new set of features and features from KDDCUP99 [15], in order to detect the fast attack from the perspective of the victim. The new selection features is depicted in table I.

*Dest_count* is derived by using the frequency episode technique, in which this technique can discover what time-based sequence of audit events frequently occur together [16].

*B. Threshold analysis*

Selecting threshold value is necessary to help the intrusion detection to make a good decision in identifying the attacker especially the fast attack [17]. It is difficult to select the suitable threshold value for differentiating between normal activity and abnormal activity in network traffic [18]. Selecting inaccurate threshold value will cause an excessive false alarm especially if the value is too low or if it is too high, it can cause the intrusion activity being considered as normal traffic. Therefore selecting an appropriated threshold is important to detect the fast attack activity. Detecting the intrusion as quickly as possible is important to prevent any losses due to security breach [19].

Threshold value can be determined by using static or dynamic techniques. Dynamic threshold technique requires training or prior knowledge of the network traffic before the threshold value can be selected [20]. This process use huge amount of information and need human expert to analyze the result, making it time consuming, thus not suitable for fast attack detection. Meanwhile, threshold value has been adopted widely in differentiating normal and abnormal network activity. Worked such as [21] used 60 connections per second from source IP address as one of the criteria to identify the intrusion, on the other hand [11] used static threshold mechanism in identifying the portscan activity while [22] used mean and standard deviation from a normal data of a host to distinguish between the normal and abnormal data and [23] generated a model of normal network traffic and then searched for the anomalous traffic that would indicate the presence of portscan.

However all these researchers only concentrates on the techniques rather than revealing the feature that influence the selection of the threshold and most of the research does not propose a proper technique to identify the threshold technique. Even though some of the researcher using the observation techniques but they do not specify a clear explanation on selecting the threshold value. To overcome this matter, this research is introduces a new technique to identify a suitable static threshold value especially for fast attack detection. Our research incorporated observation and experimental technique in finding the threshold and then verified the threshold value with SPC. Next section will discussed this technique in details

### III. METHODOLOGY

The overall threshold verification process is depicted in figure 2. The observation is done on real time network traffic from a government agencies and simulation traffic from Darpa99 [24]. The purpose of the observation is to identify the average connection made by host or hosts to single victim within one second time interval. By doing this, the connection made to single victim can be identified. The result from the observation technique will be compared to select the appropriate static threshold. Meanwhile, for the experiment technique, a small local area network (LAN) was setup. The purpose of the experiment is to identify the normal connection made by each of operating system. By doing this, the normal behavior of host in transmitting network packet to the destination host within one second time interval is identified. The result from each of the operating system will be captured and compared with each other.

Next, the appropriate threshold is selected based on the comparison from both of the observations and the experiment is compared. Subsequently, the threshold value is verified in the verification process using the SPC approach.

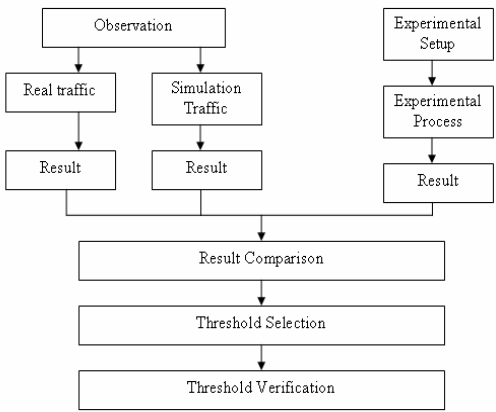

Figure 2. Threshold Selection process

#### A. Observation Technique

The technique is applied on real time network traffic. First the network traffic data is captured using *TCPdump* [25]. Next all the initial connection to each of the network traffic to the single destination host is extracted from the *TCPdump* log file. After initial connection has been extracted, the traffic will be segregated based on a well known port [26]. In this research we are interested to capture connection on port 21, 25, 53, 110, 135, 139 and 445. The reason is that these port are well known port and offer important services for the network, for example port 110 and 25 is use for email in which plays a vital role in today's business communication [27], while port 53 is important because it is the reference center for mapping IP address to DNS, if it is attacked the whole network will be in catastrophic [28]. Moreover these ports are the most popular target for attack activity especially for worm virus and port scanning [29].

After the segregation process has been done, the output from the segregation process will be fed to the feature extraction module to extract the necessary feature inside this research. Then, the result from the feature extraction module will be passed to time based module to compute the statistical value. The output for the time based module will be used to identify the mean of the connection made by host or hosts to single destination host within one second time interval using the SPSS statistical application. Meanwhile, for Darpa99 network traffic, the same process will be repeated with a slight modification. Since Darpa99 has declared the network traffic as normal network traffic, therefore, the network traffic is not segregated based on the well known port. The mean of the normal connection will be identified and both the results will be compared to identify the average to normal connection for the fast attack.

#### B. Experimental Technique

An experimental environment is designed for the purpose of identifying the ideal static threshold value in a control environment; the network setup is depicted in figure 3. One small local area network has been setup and it consists of linux centos operating system, two Windows XP Professional Service Pack 2 operating system, one fresh install Windows XP Professional Service Pack 2 operating system and one Windows Vista operating system.

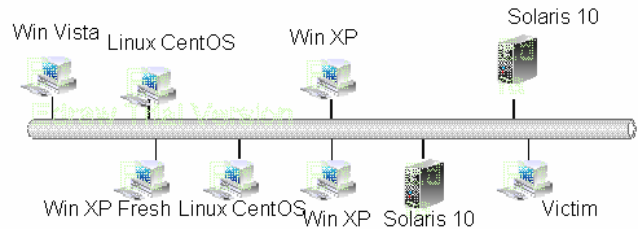

Figure 3. Network Design for Experiment

The purpose of the experiment is to identify the number of connection received by single victim from each operating system in the network. By investigating the behavior of packet transmission for each of the operating system, the normal behavior of host in transmitting network packet to single destination host within one second time interval is identified. The result may help to indicate a suitable value for the static threshold in an ideal setting.

*C. Verification Technique*

The threshold value from the observation and experimental result is then applied to the statistical process control to increase the confidence of the threshold value. The statistical process control technique is very powerful tools for detecting the changes in manufacturing process. If any event exceeds the threshold value or the control limit, it shows that the event is out of control. Out of control means that there is abnormal network traffic inside the network. Thus the administrator needs to investigate and block the network traffic before serious damages occur in the network.

Figure 4, shows the general approach proposed using statistical process control technique in validating the threshold.

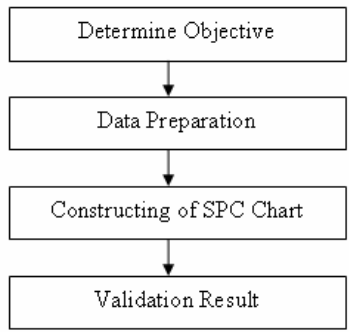

Figure 4. Statistical Process Control General Approach

The Shewhart Control Chart is used as a technique to verify the threshold. The network traffic is considered as out of control if the point falls beyond the control limit as stated using Western Electric Handbook [29]. The Western Electric Handbook suggests a set of decision rules for detecting non-random patterns on control charts. Specifically, the Western electric rules would conclude that the process is out of control if either

1. One point plot outside 3-sigma control limits.
2. Two out of three consecutive points plot beyond a 2-sigma limit.
3. Four out of five consecutive points plot at a distance of 1-sigma or beyond from the center line.
4. Eight consecutive points plot on one side of the center line.

Rule 4 is not applicable to this research because all of the normal connection will be plotted below the normal threshold and it is placed at one side of the center line. If the rule is considered in this research, then all the normal connection will be considered as a abnormal behavior which may lead to false alarm.

The general model for the control chart is introduced by Dr. Walter A. Shewart [30] Let *P* be a sample statistic that measures the quality characteristic of interest, and the mean of *P* is $\mu p$ and the standart deviation of P is $\sigma p$. The center line (*CL*), the upper control limit (*UCL*) and lower control limit (*LCL*) equation is stated in (1).

$$UCL = \mu p + k \sigma p$$
$$CL = \mu p$$
$$LCL = \mu p - k \sigma p \qquad (1)$$

Where *k* is the distance of the control limit from the center line. A common choice for *k* is 3. The k value is also used by [31] in their research in detecting the intrusion activity based on audit trail. If the reading exceeding the mean value and the *CL*, the network traffic is considered as an abnormal traffic, meaning that it is an intrusion activity.

IV. IMPLEMENTATION AND RESULT

These sections discuss the implementation and result of the observation techniques, experimental technique and result verification. Python [32] and shell scripting is used to develop the platform and the underlying operation system is Linux. The system was designed in such a way so that it can work and extract feature under real time traffic environment. The observation was done using real network traffic from a government agency and simulation data of the normal network traffic from Darpa99.

TABLE II. Average connection for site

| Port Number | Mean of connection per second | Min connection per second | Maximum connection per second |
|---|---|---|---|
| 21 | 1.16 | 1 | 3 |
| 25 | 9.57 | 1 | 70 |
| 53 | 2.91 | 1 | 9 |
| 110 | 1.07 | 1 | 2 |
| 135 | 2.83 | 1 | 46 |
| 139 | 2.83 | 1 | 46 |
| 445 | 1.14 | 1 | 3 |

*A. Observation on Site*

Based on table II, the average number of connection per second for each port falls between one to three connections per second. Only one port falls above the average of three connections per second which is port 25. From the observation, it seems that there is an abnormal behavior of network traffic from this port. There are seventy connections per second made by a single host to single victim compared to other host with only one connection per second. By referring to RFC 2821, the connection was validated as abnormal traffic. Therefore, the threshold of normal network traffic per second for the Site can be selected between one to three connections per second.

*B. Observation from Darpa99 Data*

Besides using the real traffic, Darpa99 data was also used to identify the normal connection made by hosts to single victim. Compared to the real network traffic, Darpa99 data is not divided according to the known port because the data is purely normal data. The Week 1 data from Monday to Wednesday which is stated as normal data by Darpa99 was used for the observation.

Table III, shows the summary of the mean normal connection per second using Darpa99 data. The result indicates that the host makes only one to two connections per second to a single victim. This result shows almost the same result with the real traffic as stated earlier. Due to the same result from both of the observation, it can be concluded that the normal host only makes one to three connections per second to a single victim. The value of the normal connection can be used a temporary threshold value until the comparison between the experiment technique is completed.

TABLE III. Summary of the Mean Normal Connection per Second for Darpa99 Data

| Day | Mean |
|---|---|
| Monday | 1.77 |
| Tuesday | 2.18 |
| Wednesday | 1.81 |

*C. Experimental Data*

TABLE IV. Experimental Result

| Operating System | Number of Connection Per second |
|---|---|
| Windows XP Professionals Service Pack 2 (Fresh Install) | 3 |
| Windows Vista | 3 |
| Windows XP Professionals Service Pack 2 | 3 |
| Linux CentOS 4.4 | 1 |
| Solaris 10 | 1 |

Based on the experiment result which is depicted in Table IV, the normal behavior of Windows operating system generates 3 connections per second. Meanwhile the others operating systems which are Linux and Solaris only made one connections per second. Therefore three connections per second can be considered as a normal connection.

*D. Result Verification*

The comparison between the experiment and the observation from the site and Darpa99 data shows that the mean or threshold value for a normal connection should be equal or less than 3 connections in 1 second. This value is then applied to formula 1 to produce the UCL, LCL and CL for the Shewhart Control Chart. It means that any connections that exceed the normal threshold value are suspected as attack traffic and will be examined manually to identify the accuracy of the selected threshold. Consequently, if there is no connection per second which exceeds the normal threshold value the connection is considered as normal traffic behavior.

From the Shewhart Control Chart, figure 5, we found that port 21, 110, 135 and 445 have a normal traffic connection as it traffic never exceed the threshold value. However, port 25, 53 and 139 have a data exceeding the threshold value and need further investigation. On port 25, the out of control point has been verified manually by examining the connection made by host to a victim. Based on the examination, only one host generated excessive connection to port 25. By referring to the RFC 2821 the excessive connection made by this host is considered having abnormal connection and can be classified as fast attack intrusion activity.

Whereas further examination on port 53 shows all the points which exceed the Upper Control Limit is generated from one host who make a resolution using TCP connection to one public domain name server. The host should use internal (private) DNS of the organization to make a resolution instead of public DNS which is the best practice for implementing the domain name service [33]. Therefore, this host is considered as an attacker which launches an attack to the public domain name server. Finally, the reason that causes the reading on port 139 to rise above the threshold value is due to the generation of request from the internal host which tried to expose the traffic to the external host. Based on the verification from the system administrator of the organization, port 139 has been verified as an abnormal traffic. Therefore the confirmation has been made that all of out of control connection has been compromised by worm activity.

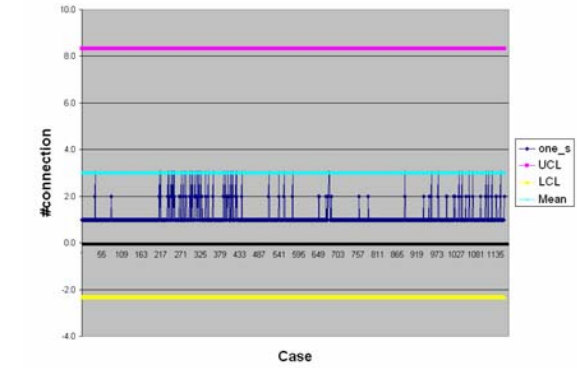

(a) Port 21

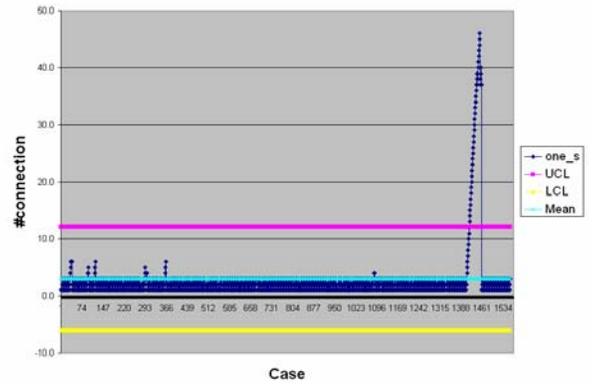

(e) Port 139

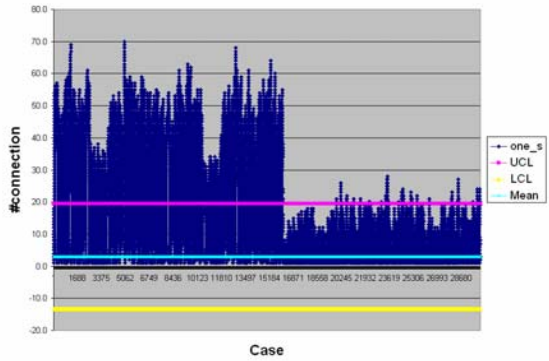

(b) Port 25

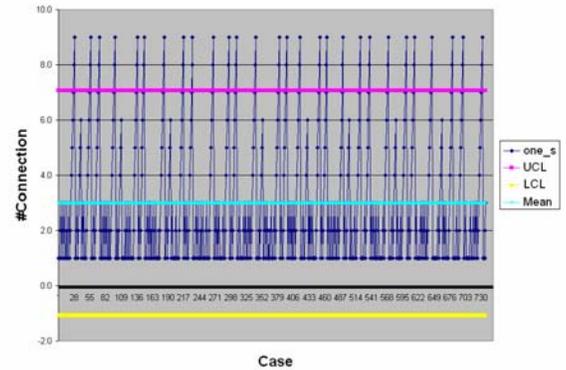

(f) Port 53

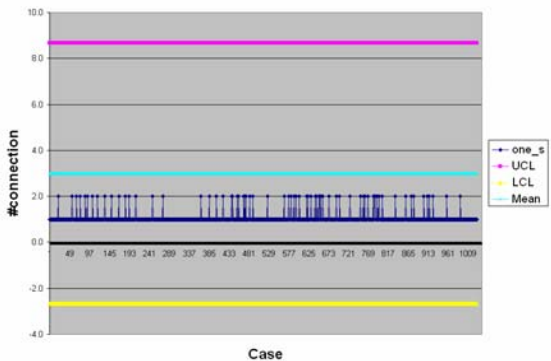

(c) Port 110

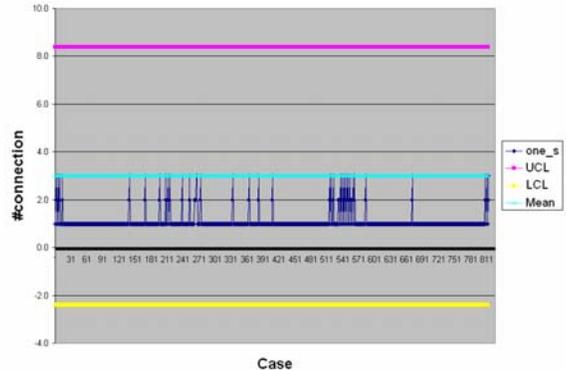

(g) Port 445

Figure 5. SPC chart for port (a) 21, (b) 25, (c) 110, (d) 135, (e) 139, (f) 53, (g) 445.

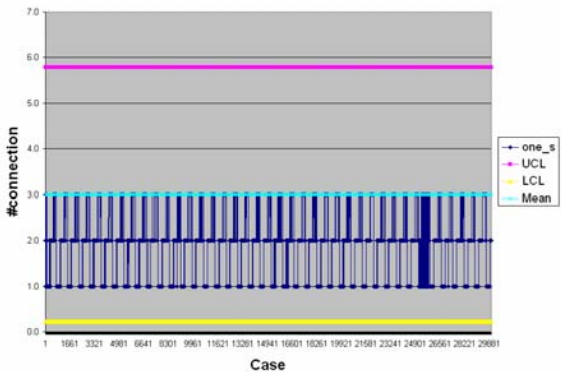

(d) Port 135

In conclusion the research found that by setting up the threshold value to 3, there are three ports which exceed the center limit of the control chart. The exceeding point from experiment indicates there is abnormal behavior of network traffic to each of the port since from the experiment, observation and darpa99 data shown that any connection more than 3 connections per second can be considered as fast attack activity. Therefore the SPC chart is suitable to validate the result of the fast attack intrusion activity and the threshold selected during the observation and experiment to distinguish between the normal and abnormal behavior of the network traffic can be used on real time environment since majority of the connection per second falls within the range of the normal behavior.

## V. CONCLUSION AND FUTURE WORKS

In this paper, we introduced a new approach in verifying threshold value for network intrusion detection system especially in detecting fast attack. The threshold value is obtained using observation and experimental technique, in which the result from both techniques is then compared and verified using statistical control process approach. By using real time network traffic data, simulation data from DARPA99 and data from the experimental setup, the research has concluded that any connection that exceed the threshold value of 3 within 1 second is considered as an abnormal activity for fast attack detection. By comparing all of these data, the value of the threshold describe the real environment of the network traffic. Thus the threshold can be used in real environment and in order to validate this result, the threshold value will be tested with different network traffic in the near future.

For the future work, we intend to use other protocol and other flag to recognize the fast attack intrusion activity. Inspecting other protocol and flag, may help to detect fast attack intrusion activities that launch using UDP or ICMP protocol. Our long term goal is to implement the system on a real time environment and introduce dynamic technique to identify the threshold for detecting the fast attack intrusion activity